\documentclass[reprint, amsmath, amssymb, aps]{revtex4-2}

\usepackage{amssymb}
\usepackage{gensymb}
\usepackage{graphicx}
\usepackage{dcolumn}
\usepackage{bm}

\usepackage{subcaption}
\usepackage{subfloat}
\usepackage{color}
\usepackage{bm}
\usepackage{ulem}
\usepackage{mhchem} 

\usepackage{tikz}
\usetikzlibrary{plotmarks}

\def\blue{\textcolor{blue}}
\def\red{\textcolor{red}}
\definecolor{asparagus}{rgb}{0.53, 0.66, 0.42}
\def\asparagus{\textcolor{asparagus}}
\def\purple{\textcolor{magenta}}

\begin{document}

\preprint{APS/123-QED}

\title{Control of Viscous Fingering Instability \\for Complex Yield-Stress Fluids using a Tapered Cell}

\author{Alban Pouplard}

\author{Peichun Amy Tsai$^\ast$}%

\affiliation{Department of Mechanical Engineering, University of Alberta,\\ Edmonton, Alberta, Canada T6G 2G8
}

\date{\today}

\begin{abstract}
Being a major limiting factor for the efficiency of various technologies, such as Enhanced Oil Recovery, the viscous fingering (or Saffman--Taylor) instability has been extensively studied, especially for simple Newtonian fluids. Here, we experimentally and theoretically demonstrate a vital control of inhibiting the viscous fingering instability for complex (yield-stress) fluids to generate a complete sweep with a flat interface. Using a rectangular tapered cell, we first experimentally show the feasibility of controlling the primary fingering instability of a complex yield-stress fluid when pushed by another less viscous one. We further develop a theoretical linear stability analysis generalized for complex fluids with a yield stress and a power-law form of viscosity to provide insights. With three complex solutions yielding different viscosity contrasts, we observe stable flat and unstable wavy interfaces depending on the gap gradient ($\alpha$) and injection flow rate ($Q$). Finally, the comparison reveals an agreeable theoretical stability criterion capable of predicting stable {\it vs.} unstable displacements for yield-stress fluids under various $\alpha$.
\end{abstract}


\maketitle


Occurring in a myriad of natural and technological processes, the displacement of a more viscous fluid by a less viscous one is common in many applications such as coating flows~\cite{Greener1980, Grillet1999},  
chromatographic separation \cite{Fernandez1996}, printing devices \cite{Pitts1961}, oil-well cementing, enhanced oil recovery (EOR) \cite{Zeng2016} or even \ce{CO2} sequestration ~\cite{Huppert2014}. Nevertheless, the unfavorable mobility or viscosity contrast in a porous medium or a narrow gap leads to an interfacial instability between the two fluids. This viscous fingering (VF) instability---manifested in a fingering shape---is responsible for a loss of efficiency in many industrial applications. This so-called Saffman--Taylor instability has been extensively studied for decades with Hele-Shaw cells, comprised of two parallel plates and used to simulate a homogeneous porous medium \cite{Saffman1958, Paterson1981, Homsy1987}. In recent years, the studies have been extended to complex fluids, often revealing more complex and broader fingers \cite{Bonn1995}. 
With yield-stress fluids, a new pattern of side-branching fingers has been observed, with smaller fingers forming on the side of the major ones \cite{Coussot1999}.

The control or inhibition of the viscous fingering is essential 
to increase the efficiency for some of the aforementioned industrial processes. Several strategies have been studied recently to suppress the fingering instability for simple Newtonian fluids, including the use of a time-dependent flow rate \cite{Cardoso1995, Dias2012, Fontana2014},
an elastic membrane as the top plate of a Hele-Shaw cell \cite{Pihler2012, Pihler2013, Housseiny2013}, or a tapered cell \cite{Housseiny2012, Bongrand2018}. However, such control has never been reported for complex fluids, which are omnipresent in natural and industrial settings. It has been shown recently that for a yield-stress fluid, the side-finger appearance can be affected by using a converging or diverging gap \cite{Eslami2020_1}, but the complete annihilation of the perturbation and total sweep efficiency have never been achieved.

\begin{figure*}[htb!]
    \begin{center}
    \includegraphics[width= 6in]{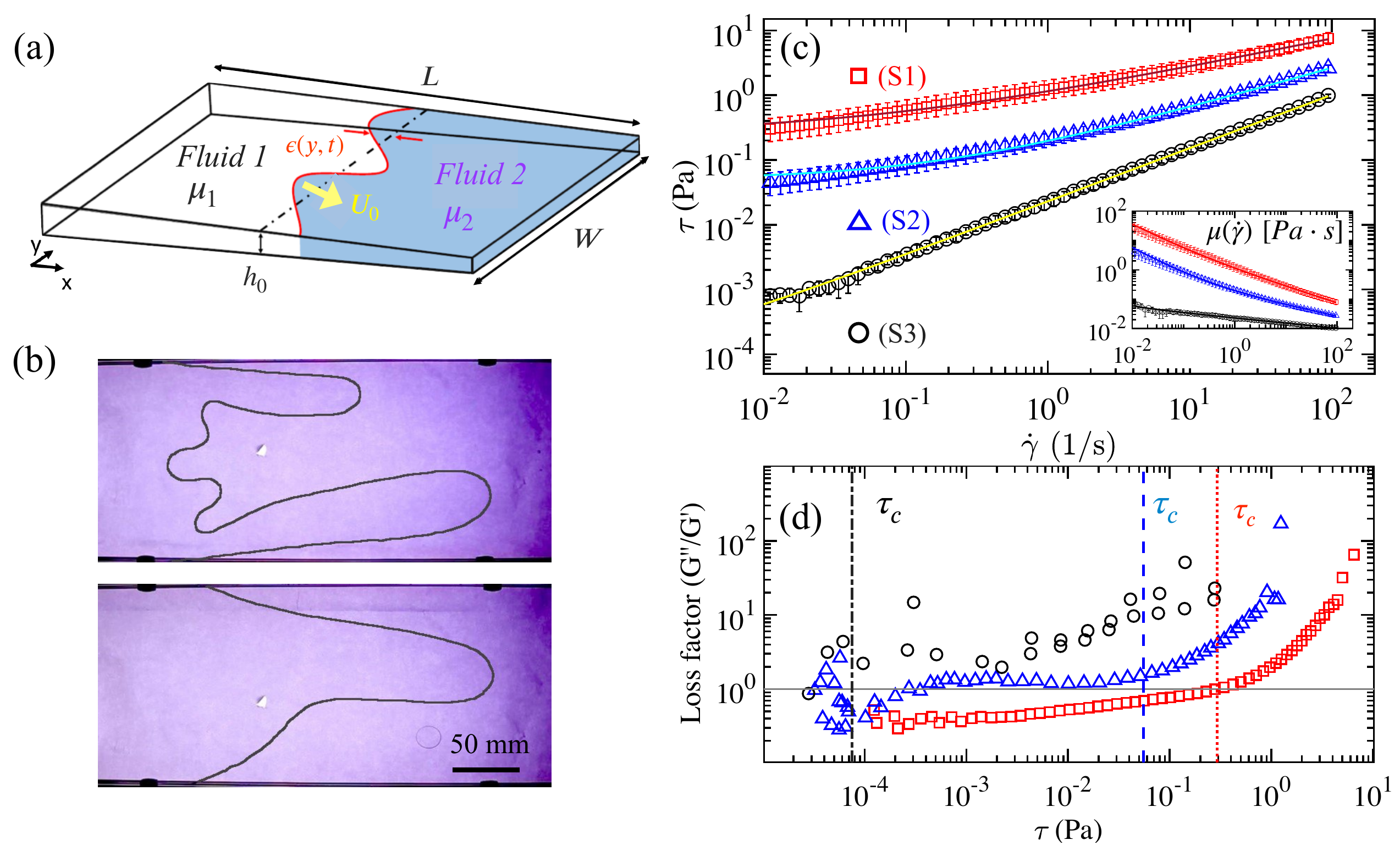}
    \end{center}
    \vspace{-0.2in}
   \protect{\caption{(a) Schematics of a fluid-fluid displacement experiment where one complex fluid of high viscosity ($\mu_2$) is invaded by another immiscible one of smaller viscosity ($\mu_1$).  (b) Experimental snapshots of complex viscous fingering produced by a PAA solution (S2) displaced by a gas at $Q = 0.2$~slpm and $Q = 0.02$~slpm in a flat Hele$-$Shaw cell. (c) Flow curves of shear stress ($\tau$) and viscosity ($\mu$), depending on the shear rate ($\dot{\gamma}$) for the three complex fluids: (S1, \red{$\square$}), (S2, \blue{$\bigtriangleup$}), and (S3, {\LARGE$\circ$}). The lines correspond to the Herschel-Bulkley model [Eq.~\eqref{eq1_HB}] fitted to data. (d) The measurements of loss factor---the ratio of the loss modulus ($G''$) to the storage modulus ($G'$)---with respect to the shear stress ($\tau$) using oscillation amplitude sweep tests. The vertical lines (from right to left) in (d) represent the yield-stress ($\tau_c$) values for the complex fluids, (S1, \red{$\square$}), (S2, \blue{$\bigtriangleup$}), and (S3, {\LARGE$\circ$}), respectively.}
    \label{Fig1}}
\end{figure*}

In this Letter, using a rectangular tapered cell, we examine the possibility of inhibiting the primary VF instability of complex, yield stress fluids in a narrow confinement (see Fig.~\ref{Fig1}). First, we investigate the effects of flow rate ($Q$) and depth gradient ($\alpha$) on the primary VF instability with experiment of a simple gas pushing a yield-stress, shear-thinning fluid. Second, we theoretically derive a linear stability criterion generalized for two complex yield-stress fluids pushing one another in a gap-converging cell. Finally, using the experimental results of interface velocity and gap thickness, good agreement between the theoretical stability criterion and the experimental results is found.

Experimentally, three different aqueous solutions of PolyAcrylic Acid solution (PAA, SigmaAldrich, $M_w \approx 1,250,000$) were used as a wetting yield-stress fluid to investigate complex fluids' viscous fingering. We initially fill in one complex PAA solution in a rectangular Hele--Shaw or tapered cell and subsequently inject a gas (Nitrogen, viscosity $\mu_1 = 1.76 \times 10^{-5}$ Pa$\cdot$s at $20~\degree$ C) as an invading fluid (see Fig.~\ref{Fig1}a). The gas is injected at a constant $Q$, ranging from $0.02$ to $2$~slpm (Alicat) with $1$~ml/min accuracy.  
The three PAA solutions prepared have different viscosities ($\mu_2$) and, hence, mobility contrasts, ${\cal M} = \mu_2/\mu_1$, spanning $4.47\times10^{4} - 1.68\times10^{6}$, $1.55 \times10^{4} - 2.64\times10^{5}$, and $5.83\times10^{2} - 3.13 \times 10^{3}$ for the complex fluids (S1), (S2), and (S3), respectively. The procedures of preparing the solutions include slowly adding the polymer powder to water and, subsequently, stirring the mixture at high speed for 1~hr. The mixture generates an acid solution that can be neutralized using a basic solution. As such, sodium hydroxide (NaOH) is added and then stirred for 10 hours at medium speed. Finally, 
the agitated solution is to rest for a day before taking any rheological measurements.

The rheological data of the three complex solutions with a rheometer (AntonPaar MCR302) are presented in Fig.~\ref{Fig1}c. Neglecting the elastic properties, the common rheological model of the Herschel$-$Bulkley (HB) model \cite{Herschel1926} reveals an excellent fit of the shear stress ($\tau$) varying with shear rate ($\dot{\gamma}$):
\begin{align}
\tau &= \tau_{c} + \kappa \dot{\gamma}^{n},\label{eq1_HB}
\end{align}
with $\tau_{c}$, $\kappa$, and $n$ corresponding to the yield stress, the consistency index, and the power-law index, respectively. The fluid's viscosity ($\mu$) varying with $\dot{\gamma}$ is well described by the corresponding HB model with $\mu = \frac{\tau_{c}}{\dot{\gamma}} + \kappa {\dot{\gamma}}^{n-1}$. Listed in Table~\ref{table1} are the best nonlinear-fits of the flow curve of $\tau = f(\dot{\gamma})$ for $\dot{\gamma}$ between $0.014$ and $93$~s$^{-1}$. The solutions are shear-thinning, with decreasing $\mu$ with increasing $\dot{\gamma}$, i.e., $n < 1$, as shown by $\mu(\dot{\gamma})$ in Fig.~\ref{Fig1}c inset.

\begin{table}[htb!]
\caption{Rheological parameters found for the three complex, yield-stress fluids with the HB model [Eq.~\eqref{eq1_HB}].
}
\label{table1}
\begin{ruledtabular}
\begin{tabular}{l c c c c c}
\textrm{Yield-stress} & 
\textrm{PAA} & 
\textrm{NaOH}&
\textrm{$\tau_{c}$}&
\textrm{$\kappa$}&
\textrm{$n$}\\
\textrm{solution} &
\textrm{(wt~$\%$)} &
\textrm{(wt~$\%$)} &
\textrm{(Pa)} &
\textrm{(Pa.s$^n$)} \\
\colrule
(S1) & 0.025 & 0.0106 & 0.2940 & 0.8641 & 0.4623 \\
(S2) & 0.10 & 0 & 0.0552 & 0.1442 & 0.6286 \\
(S3) & 0.015 & 0.008 & 7.51e-5 & 0.0231 & 0.8208 \\
\end{tabular}
\end{ruledtabular}
\end{table}

Oscillation amplitude sweep tests at constant frequency ($\hat{\omega} = 1$~rad/s) are perform to validate negligible elasticity of the complex fluids. Fig.~\ref{Fig1}d shows the results of the loss factor, i.e., the ratio of the loss modulus $G''$ (representing the viscous properties) to the storage modulus $G'$ (corresponding to the fluid elasticity) varying with $\tau$. The viscous behavior prevails when $G''/G' >1$, whereas the elasticity is dominant when the ratio $< 1$. The vertical lines (from right to left) in Fig.~\ref{Fig1}d represent the values of the yield-stress ($\tau_c$) for the complex fluids (S1), (S2), and (S3). We only focus on the flowing regime, i.e, $\tau > \tau_c$. Whenever this condition is fulfilled, the loss factor $\gtrsim$ 1, meaning the viscous component prevails. We can neglect the elastic effects for (S1) and (S2) from this observation. The loss factor for the fluid (S3) is always $\gtrsim 1$, so its elasticity can also be neglected.

\begin{figure*} [htb!]
    \begin{center}
    \includegraphics[width=7in]{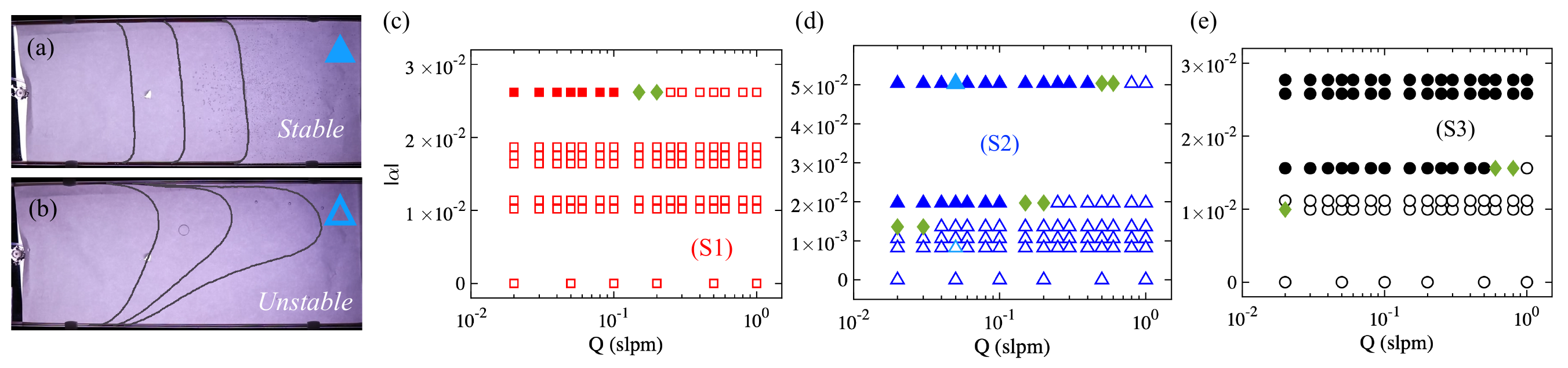}
    \end{center}
    \vspace{-0.2in}
    \caption{{\bf Control of complex viscous fingering} using a rectangular tapered cell with a linearly varying gap thickness ($h = h_e + \alpha x$), demonstrated by the overlay of experimental snapshots from a stable and an unstable displacement in (a) and (b), respectively.
    (a) the stable displacement is obtained when nitrogen is pushing complex fluid (S2) in a tapered cell with $h_e = 23.24$~mm, $\alpha$ = -5.03 $\times 10^{-2}$ and $Q = 0.05$~slpm. The time step between the snapshots is $\delta t = 100$~s. (b) the unstable displacement is observed when a gas is pushing (S2) in another taper with $h_e = 3.76$~mm , $\alpha= -8.20\times 10^{-3}$, $Q = 0.05$~slpm, and $\delta t=20$~s. (c), (d), and (e) are experimental stability diagrams for fluids (S1), (S2) and (S3), respectively, with stable uniform (filled symbols; \red{$\blacksquare$}, {\large\blue{$\blacktriangle$}} and {\LARGE $\bullet$}) {\it vs.} fingering/wavy unstable interfaces (open symbols; \red{$\square$}, \blue{$\bigtriangleup$} and {\LARGE$\circ$}) under various values of $Q$ and $\alpha$.  Finally, green diamonds (\asparagus{$\blacklozenge$}) correspond to the transitional state when the interface starts to develop a wavy profile.
    \label{Fig2}}
\end{figure*}

In the experiments, a rectangular cell consisting of two glass plates with a converging gap is used to control the VF instability. The cell length and width are $L=458$~mm and $W=153$~mm, respectively. The snapshots are captured via bottom view with a camera (Canon) at 30 fps. We use ImageJ and Matlab to analyze the images and track the interface position. The interface velocity ($U_0$) is calculated by deriving the interface position over a short time. The values of $\alpha$ and the interface's position and gap thickness ($h_0$) are determined from image analysis using the values of $U_0$ and $Q$.


We carry out approximately 550 experiments using rectangular flat and tapered cells. With the flat Hele--Shaw cell of a constant gap-thickness, we always observe an unstable interface with one or multiple fingers of the complex fluid developing at the interface, consistent with previous similar observations using more viscous complex fluids \cite{Eslami2017}. 
Shown in Fig.~\ref{Fig1}b are two representative snapshots obtained in our experiments when (S2) is pushed by nitrogen at $Q = 0.2$~slpm and $Q = 0.02$~slpm, respectively. Furthermore, for the two more viscous fluids (S1) and (S2) and with greater $Q$, the formation of small fingers is observed on the side of the major one, corresponding to the side-branching regime \cite{Jirsaraei2005} or the elasto-inertial regime \cite{Eslami2017}. The side-fingering pattern is observed when $Q$ is above a critical value, $Q_s$. $Q_s = 0.5$ and $1$~slpm for (S1) and (S2), respectively.

By contrast, we do not observe the side-branching patterns when the converging cells are used. Instead, the interfacial patterns observed resemble either the usual viscous fingers of Newtonian fluids or a stable interface, occurring at a low flow rate depending on ${\cal M}$ and $\alpha$. The critical flow rates, below which we observe a stable displacement, for each complex fluid and $\alpha$ are summarized in Fig.~\ref{Fig2}c-e. The stable displacement is characterized by a complete sweep efficiency, as shown by Fig.~\ref{Fig2}a. 
Such inhibition of the primary VF instability is only possible for suitable rheological and flow parameters, as shown by the stability phase diagrams in Fig.~\ref{Fig2}c-e for the complex fluids (S1) to (S3), respectively. The essential experimental finding is that we observe stable interfaces for all the fluids (S1) to (S3) for lower $Q$ as the gap-gradient value ($|\alpha|$) increases. The stable interfaces occur at low $Q$ for a constant gap gradient.

In Fig.~\ref{Fig2}c-e, we differentiate three types of experimental results: first, uniform and stable  displacements represented by filled symbols (S1, \red{$\blacksquare$}), (S2, {\large\blue{$\blacktriangle$}}), and (S3, {\Large $\bullet$}); second, unstable displacements with fingering or wavy interfaces by open symbols: (S1, \red{$\square$}), (S2, \blue{$\bigtriangleup$}) and (S3, {\LARGE$\circ$}); third, the transitional state when the interface starts to develop a wavy profile (\asparagus{$\blacklozenge$} for all the fluids).  The contrast between the stability diagrams of the three different complex solutions highlight not only the complexity of controlling complex VF but also the importance of rheologies via $\tau_c$, $\kappa$, $n$, and local $\dot{\gamma}$.

To obtain a better understanding of the important parameters to control the VF instability of complex fluids, we carry out a linear stability analysis using two yield-stress, complex fluids (Fluid 1 pushing Fluid 2) of viscosity $\mu_1$ and $\mu_2$, respectively, in a rectangular, tapered cell of a length, $L$, and width, $W$ (see Fig.~\ref{Fig1}a). A constant $\alpha$ produces a linearly varying height ($h$) between the two plates of the cell. We consider a lubrication flow confined in the thin gap, for which the height varies linearly in the $x$ direction as $h(x) = h_{0} + \alpha x$. $h_{0}$ denotes the gap thickness at the two fluids' interface, located at $x=0$.

In the derivation, we use the effective Darcy's law replacing the constant viscosity, $\mu$, by the effective shear-dependent viscosity, $\mu_\text{eff}$. This approach has been used in modeling the problems of non-Newtonian fluid-fluid displacement in a flat Hele--Shaw cell \cite{Bonn1995}, but here we extend the model to a tapered geometry. 
Neglecting the fluids' elastic properties \cite{Coussot1999}, the governing equations are the 2D depth-average Darcy and continuity equations:
\begin{align}
{\bf U}_j = -\frac{h^2}{12\mu_{\text{eff}j}} \vec{\nabla} P_j~,~~~
\nabla \cdot \left(h {\bf U}_j \right) = 0 . \label{eq.2.appxA}
\end{align}
${\bf U}_j(x,y) = (u_{xj} , u_{yj})$ and $P_j(x,y)$ are the depth-average velocity and pressure fields of the fluid indexed $j$, respectively. $j$ represents the two complex fluids during the displacement process; $j = 1$ (2) denotes the pushing (displaced) complex fluid. The complex fluid's viscosity ($\mu_{\text{eff} j}$) is modeled using the HB law [Eq.~\eqref{eq1_HB}] for yield-stress fluids \cite{Herschel1926}, with the local shear rate $\dot{\gamma} =  \frac{u_{xj}}{h}$, and expressed as
$\mu_{\text{eff} j} = \frac{\tau_{cj}}{\overset{.}{\gamma}} + \kappa_j {\overset{.}{\gamma}}^{n_j-1}$.
The dimensionless Bingham number is defined as the ratio of the yield to viscous stress: $Bn_j= \frac{\tau_{cj}}{\kappa_j \left(\frac{u_{xj}}{h} \right)^{n_j}}$. Assuming a small ratio of gap change: $\frac{\alpha  x}{h_0} \ll 1$, we can linearize the expression of the gap thickness as $h = h_0 \left(1 +\frac{\alpha x}{h_{0}} \right)$ and neglect the higher-order terms of $O \left(\alpha^2 \right)$. The depth-average continuity in Eq.~\eqref{eq.2.appxA} can be expressed using the pressure field ($P_j$) and further simplified. 
For $n_j = 1$ and $\tau_{cj} = 0$, we recover the following equation: $\frac{\partial^2 P_j}{\partial x^2} + \frac{3 \alpha}{h} \frac{\partial P_j}{\partial x} + \frac{\partial^2 P_j}{\partial y^2} = 0$, which is found by Al-Housseiny and Stone for the simple Newtonian counterpart \cite{Stone2013}.

\begin{figure*} [htb!]
    \centering
    \includegraphics[width=6in]{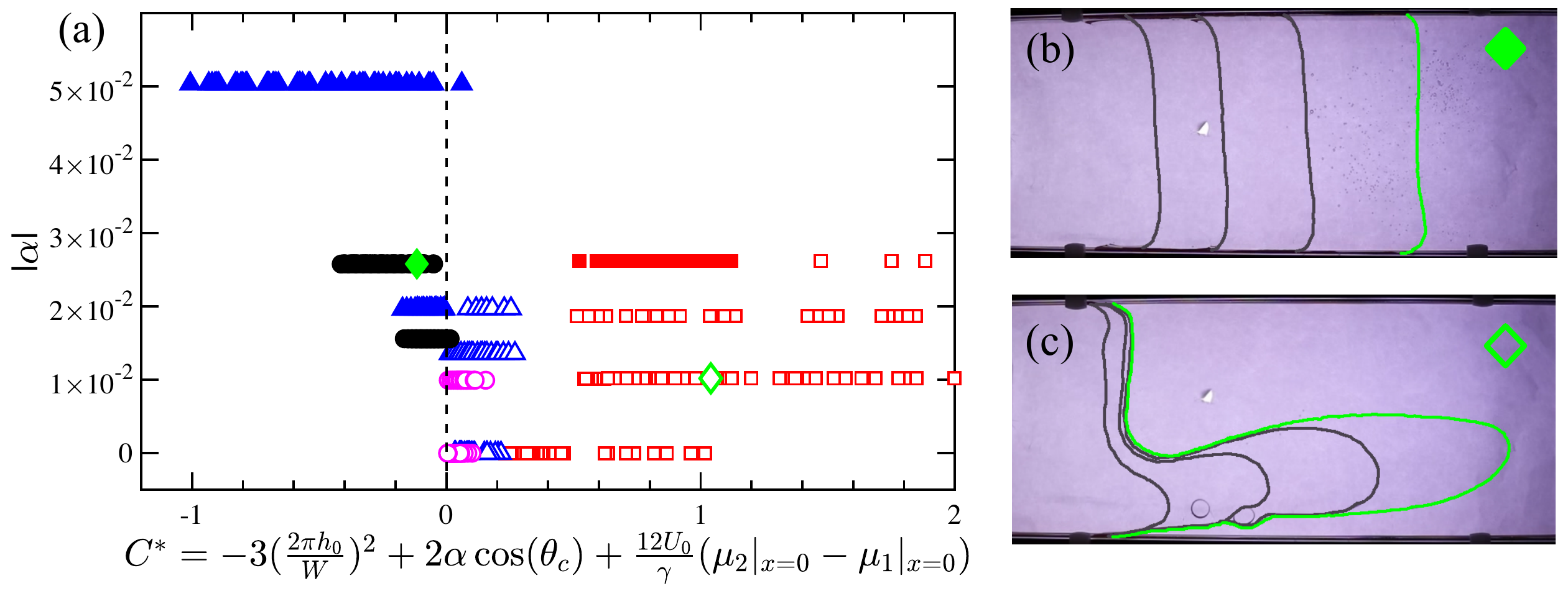}
    \vspace{-0.1in}
    \caption{{\bf Comparison between experimental and theoretical results:} (a) Dimensionless term $C^\ast$ [Eq.~\eqref{eq.23.appxA}] for different experiments performed with different gap gradient ($\alpha$). The stable (unstable) experiments with solution (S1), (S2) and (S3) are represented by filled (open) symbols, \red{$\blacksquare$}, {\large\blue{$\blacktriangle$}} and {\LARGE $\bullet$} (\red{$\square$}, \blue{$\bigtriangleup$} and \purple{\LARGE$\circ$}), respectively. (b) and (c) show the overlay of experimental snapshots with one stable and one unstable displacement (and $\delta t =150$s and $\delta t =4$s), respectively. The interfaces highlighted in green correspond to the two symbols in (a).} 
    \label{Fig3}
\end{figure*}

In the linear stability analysis, the pressure is expressed as the addition of the base state and perturbation: $P_j (x,y,t) = f_j (x) + g_{kj} (x) \epsilon (y,t)$, where $f_j(x)$ corresponds to the base-state pressure field when the interface is stable and independent of $y$. The term of $g_{kj} (x) \epsilon$ represents the perturbation that propagates along the interface, with the perturbation $\epsilon (y,t) = \epsilon_0 \exp{\left( i k y+ \sigma t \right)}$ depending on the perturbation's wavenumber ($k$) and growth rate ($\sigma$). The kinematic boundary condition is applied to ensure the two fluids move at the same interfacial velocity. The capillary pressure jump at the interface is described by the Young--Laplace equation due to interfacial tension. We assume a small ratio of gap change and a small Bingham number $Bn_j \ll 1$, i.e., the fluid yield stress is negligible compared to the viscous stress.
By substituting the expression of the linearized pressure into the pressure jump condition and removing all the base state components,
the Laplace pressure Eq. transforms into the following dimensionless dispersion-relation with the dimensionless growth rate defined as $\overline{\sigma} = \frac{\sigma {h_0}}{U_0}$, and the dimensionless wavenumber, $\overline{k} = h_0 k$ (See Supplementary Information for details):
\begin{widetext}
\begin{equation} \label{eq.21.appxA}
\begin{split}
\frac{12 \overline{\sigma} h_0 \left(\kappa_1 \sqrt{n_1} \left(\frac{U_0}{h_0} \right)^{n_1} + \kappa_2 \sqrt{n_2} \left(\frac{U_0}{h_0} \right)^{n_2} \right)}{\gamma} & = -\frac{12 \alpha h_0}{\gamma} \left(\kappa_1 \sqrt{n_1} \left(\frac{U_0}{h_0} \right)^{n_1} + \kappa_2 \sqrt{n_2} \left(\frac{U_0}{h_0} \right)^{n_2} \right) \\
& + \overline{k} \left(2 \alpha \cos{\theta_c} +\frac{12 U_0}{\gamma} \left(\mu_2|_{x=0} - \mu_1|_{x=0} \right) \right) -\overline{k}^3.
\end{split}
\end{equation}
\end{widetext}

To find the wavenumber or wavelength at the maximum growth, we take the derivative of Eq.~\eqref{eq.21.appxA} w.r.t. $\overline{k}$. Consequently, by setting $\frac{ \partial \overline{\sigma}}{\partial \overline{k}} = 0$, we find the wavenumber of maximum growth ($\bar{k}_{max}$) and the corresponding wavelength of maximum growth, $\lambda_{max} = \frac{2 \pi h_0}{\bar{k}_{max}}$:
\begin{equation*}
\lambda_{max} = 2 \pi h_0 \left(\frac{3}{2 \alpha \cos{\theta_c} + \frac{12 U_0}{\gamma} \left(\mu_2|_{x=0} - \mu_1|_{x=0} \right)} \right)^{\frac{1}{2}}.
\end{equation*}

The viscous fingering instability of complex fluids will be apparent when $\lambda_{max} < W$ (i.e., the cell width), leading to the following criterion for visible fingering:
\begin{equation}\label{eq.23.appxA}
- 3 \left(\frac{2 \pi h_0}{W} \right)^2 + 2 \alpha \cos{\theta_c} + \frac{12 U_0}{\gamma} \left(\mu_2|_{x=0} - \mu_1|_{x=0} \right) > 0.
\end{equation}
This criterion depends on the local rheological properties of the fluids ($\kappa_j, n_{j}, \tau_{cj}$), the gap gradient ($\alpha$) as well as the velocity ($U_0$) and gap thickness ($h_0$) at the interface, in addition to various factors such as the wetting angle ($\theta_c$) and the interfacial tension ($\gamma$) between the pushing and the driven fluid. We further compare our experimental results with this theoretical criterion Eq.~\eqref{eq.23.appxA} for a visible complex viscous finger.
Using the experimental values of $U_0$, $h_0$, and $\alpha$, we plot our stability criterion [Eq.~\eqref{eq.23.appxA}] in Fig.~\ref{Fig3}a. Here, we consider a stable displacement when the fluid-fluid interface is uniform and stable throughout the entire experiment.

In accordance with our theory, a clear separation between stable (\red{$\blacksquare$}, {\large\blue{$\blacktriangle$}} and {\Large $\bullet$}) and unstable displacements (\red{$\square$}, \blue{$\bigtriangleup$} and {\LARGE\purple{$\circ$}}) for a criterion around 0 is shown in Fig.~\ref{Fig3}a. While good agreements are found for (S1) and (S2), but for the more viscous fluid (S1) we observe stable interfaces that are deviated from the expected chart. This deviation may be explained by the few assumptions made. The impact of the gravity and the elastic properties have been ignored in our theoretical derivation. Moreover, whenever $|\alpha|$ becomes large, the assumptions of the small ratio of gap change ($ \frac{\alpha x}{h_0} \ll 1$) as well as the characteristic length scale over which the depth varies much larger than that of the perturbation length scale ($\frac{k h_0}{\alpha} \gg 1$) become unjustified. Finally, we neglect the yield stress with respect to the viscous stress and use a constant static contact angle. Further theoretical investigations are needed to mitigate these assumptions, aiming at better agreement with experiments especially for very viscous complex fluids.

In summary, we demonstrate experimentally and theoretically a valuable and powerful method of inhibiting the VF instability for complex yield-stress fluids by using a converging tapered rectangular cell. We perform a linear stability analysis using an effective Darcy's law and derive a stability criterion concerning the perturbation wavelength. For the Newtonian fluids, the stability of the interface is determined by four parameters, namely the gap gradient, contact angle, viscosity contrast, and Capillary number ($\alpha$, $\theta_c$, $\lambda$, and $Ca$, respectively). In contrast, for complex yield-stress fluids, in addition to the above four parameters, the cell width  ($W$), the interface's velocity and gap thickness ($U_0$ and $h_0$), as well as the HB coefficients ($\tau_c$, $\kappa$, and $n$) determining the local viscosity influence the VF stability through Eq.~\eqref{eq.23.appxA}.

For various $\alpha$ ranging from $-5.03 \times 10^{-2}$ to $0$, we observe a transition from stable and unstable interfaces at the stability criteria by $C^\ast$ in Eq.~\eqref{eq.23.appxA} close to 0, as shown in Fig.~\ref{Fig3}a. However, for the more viscous fluid (S1), we observe a discrepancy when $|\alpha|$ is greater. This deviation may stem from some key assumptions made, for instance, the impacts of the gravity, elastic properties of the fluids, and the Bingham number have all been neglected. Moreover, we made assumptions using a static contact angle and a small gap gradient ($\alpha$). Nevertheless, we experimentally demonstrate a successful control of complex yield-stress VF instability using a tapered cell under various values of $\alpha$ and $Q$. These insightful experimental and theoretical results can lead to a wide variety of applications for controlling interfacial patterns for complex yield-stress fluids during fluid-fluid displacement processes in microfluidics, narrow confinements, and porous media, by selecting suitable rheological parameters, $U_0$, $h_0$, and $\alpha$.  

\begin{acknowledgments}
The authors gratefully acknowledge the funding support from the Natural Sciences and Engineering Research Council of Canada (NSERC) Discovery grant (RGPIN-2020-05511). P.A.T. holds a Canada Research Chair in Fluids and Interfaces (CRC TIER2 233147). This research was undertaken, in part, thanks to funding from the Canada Research Chairs (CRC) Program.
\end{acknowledgments}

$^\ast$ Email: peichun.amy.tsai@ualberta.ca

\bibliography{Ref}

\end{document}